\pdfoutput=1

\documentclass[aps,prb,twocolumn,amsfonts,amssymb,superscriptaddress,longbibliography]{revtex4-2}
\usepackage{graphicx}
\usepackage{dcolumn}
\usepackage{bm}
\usepackage{bbm,bbold}
\usepackage{color}
\usepackage{xcolor, soul}
\sethlcolor{green}
\usepackage{amsfonts}
\usepackage{amsmath}
\usepackage{mdframed}
\usepackage[normalem]{ulem}
\usepackage{mathrsfs}   
\usepackage[none]{hyphenat}
\usepackage{subfigure}
\usepackage{float}
\usepackage{cancel}
\usepackage{physics}
\usepackage[colorlinks=true,citecolor=blue]{hyperref}
\hypersetup{colorlinks=true,citecolor=blue,linkcolor=blue,urlcolor=blue}

\DeclareMathAlphabet\mathbfcal{OMS}{cmsy}{b}{n}

\allowdisplaybreaks 

\begin{document}
\title{Nonlinear thermoelectric probes of anomalous electron lifetimes in topological Fermi liquids}
\author{Johannes Hofmann}
\email[Corresponding author: ]{johannes.hofmann@physics.gu.se}
\affiliation{Department of Physics, Gothenburg University, 41296 Gothenburg, Sweden}
\affiliation{Nordita, Stockholm University and KTH Royal Institute of Technology, 10691 Stockholm, Sweden}
\author{Habib Rostami}
\email[Corresponding author: ]{hr745@bath.ac.uk}
\affiliation{Department of Physics, University of Bath, Claverton Down, Bath BA2 7AY, United Kingdom}

\date{\today}
\begin{abstract}
In two-dimensional Fermi liquids (FLs), odd-parity Fermi surface deformations have anomalously slow relaxation rates that are suppressed as $T^4$ with temperature $T$, distinct from the standard FL $T^2$ scaling. We demonstrate here that these long-lived modes, which are often hidden in linear response, have a significant impact on nonlinear transport by establishing a direct proportionality of nonlinear thermoelectric currents to the anomalously large relaxation time. These currents exist in topological time-reversal invariant FLs,  and their magnitude is characterized by topological heat capacitance terms that we refer to as the {\em Berry curvature capacity} and the {\em velocity-curvature capacity}. We quantify the effect in bismuth telluride, which is an efficient thermoelectric and a topological insulator with a hexagonal Fermi surface. The proposed field-induced thermoelectric currents are well within the range of current experiments. Our findings demonstrate the potential to explore topological and many-body effects in FLs through the nonlinear thermoelectric response, urging further experimental studies. 
\end{abstract}
\maketitle

{\em Introduction.} Recent advances in the fabrication of ultraclean two-dimensional (2D) materials create interaction-dominated electron gases~\cite{dejong95,buhmann02,bandurin16,crossno16,moll16,nam17,krishnakumar17,gooth18,gusev18,bandurin18,braem18,berdyugin19,gallagher19,sulpizio19}. As the temperature is increased to a sizable fraction of the Fermi temperature $T_{\rm F}$ in these materials, there is a crossover from a ballistic or diffusive transport regime to a hydrodynamic regime, where electron-electron interactions dominate over scattering from disorder and phonons. The change in the relaxation mechanism has a profound impact on the transport properties: For example, the characteristic relaxation time $\tau$ in Ohm's law \mbox{$\sigma = n e^2 \tau/m^*$} diverges since binary collisions cannot relax the total charge current. Transport is instead set by the shear viscosity \mbox{$\eta \sim \tau_{\rm FL}$}, which diverges at low temperatures with the Fermi liquid (FL) lifetime as \mbox{$\tau_{\rm FL} \sim (T_{\rm F}/T)^2$}~\cite{baym04,giuliani05,pines18,giuliani82,zheng96,li13,dassarma21,gran23}. This characteristic quadratic temperature dependence reflects a phase-space constraint where only quasiparticles close to the Fermi surface interact, and a common expectation is that this FL scale sets the temperature dependence of all relaxation times (excluding conserved modes like the current mode).

\begin{figure}[h!]
    \centering
    \includegraphics[width=5.8cm]{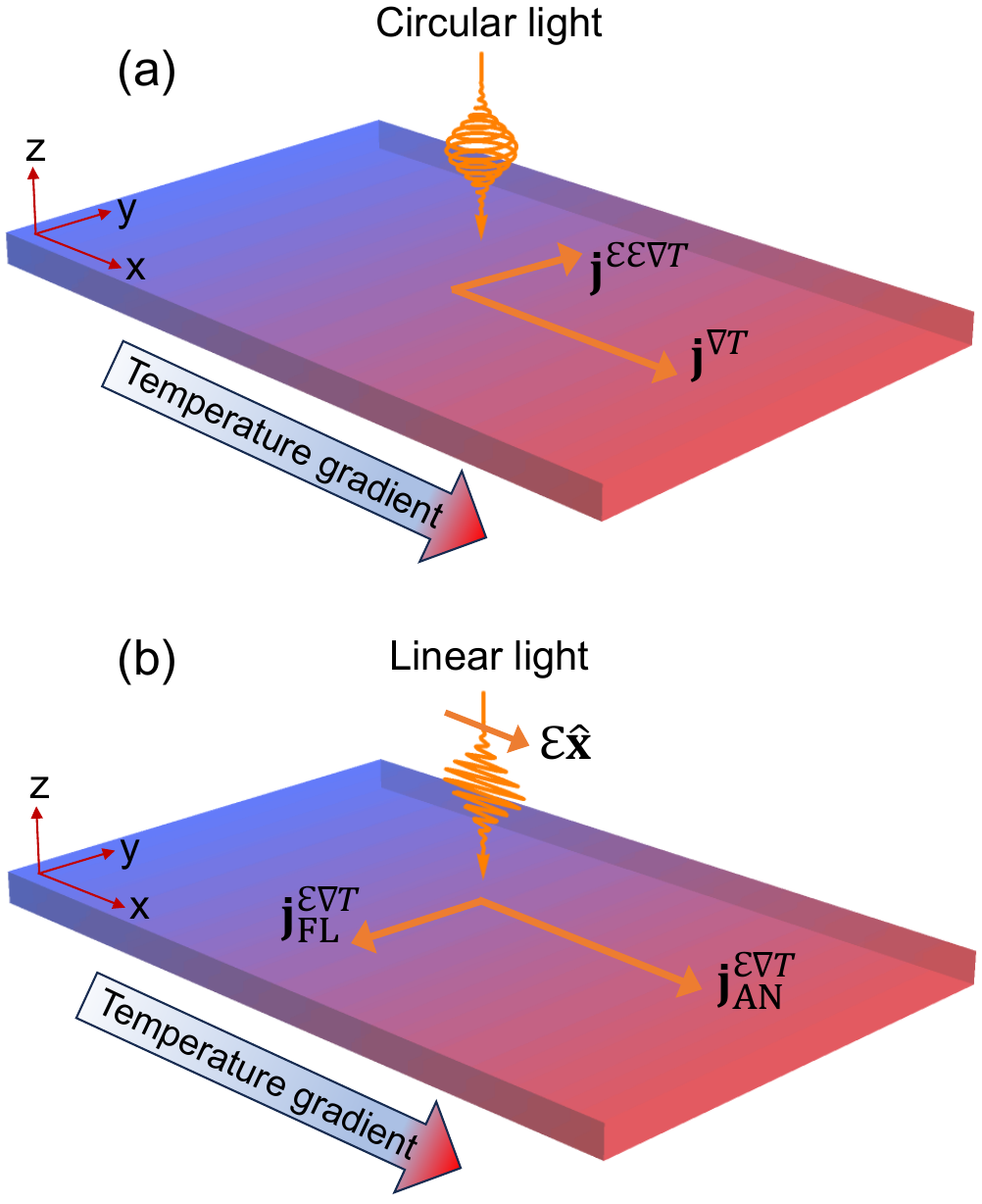}
    \caption{Illustration of field-induced thermoelectric currents. (a) Nonlinear thermoelectric current ${\bf j}^{{\cal EE}\nabla T}$ due to circular light, Eq.~\eqref{eq:jEEdT}, characterized by the anomalous relaxation time $\tau_{\rm AN}$ and Berry curvature capacity $C_\Omega$, and conventional thermoelectric current ${\bf j}^{\nabla T}$. (b) Nonlinear thermoelectric current induced by linear polarized light, characterized by the velocity-curvature capacity ${\bf C}_{\bf v}$. ${\bf j}^{{\cal E}\nabla T}_{\rm AN}$ and ${\bf j}^{{\cal E}\nabla T}_{\rm FL}$ are the orthogonal contributions proportional to $\tau_{\rm AN}$ and $\tau_{\rm FL}$, Eq.~\eqref{eq:jEdT}. We assume a temperature gradient and light polarization along the $x$ direction, normal to the mirror plane of the system (\mbox{$x\to -x$}). 
    }
    \label{fig1}
\end{figure}

However, the simple quadratic-in-temperature FL scaling only applies to a restricted class of collective deformations that have even parity. By contrast, odd-parity deformations are predicted to have anomalously long lifetimes that diverge as a quartic function of temperature \mbox{$\tau_{\rm AN} \sim (T_F/T)^4$}. Here, odd and even parity refers to the components of the electron distribution function that are odd or even under momentum inversion, \mbox{${\bf p}\to -{\bf p}$}. This odd-even anomaly had been noted previously in the literature~\cite{laikhtman92,nilsson05, gurzhi95} but has only recently been quantified, analytically at low temperatures by~\textcite{ledwith17,ledwith19} and numerically by~\textcite{hofmann23,nilsson24} for a parabolic dispersion. 
Generally, the odd-even effect requires time-reversal symmetry, \mbox{$\varepsilon(\mathbf{p}) = \varepsilon(-\mathbf{p})$}, ensuring no mixing of opposite-parity deformations. Furthermore, it is independent of the specific electron interaction since the anomalous relaxation occurs via soft small-angle scattering \cite{ledwith17}. 
However, these exotic FL features are still waiting for an experimental confirmation. Authors of previous studies on nontopological materials have suggested that odd-parity dynamics primarily impacts the transverse response, introducing additional transverse diffusive modes~\cite{hofmann22} and higher-derivative corrections to shear viscosity~\cite{ledwith19b}, but a practical measurement scheme that probes long-lived odd-parity modes is currently lacking. 

In this letter, we show that nonlinear thermoelectric transport in anisotropic and topological Fermi liquids is sensitive to electron interactions and is moreover significantly enhanced by the anomalously long-lived odd-parity features. The key element of our proposal is that breaking spatial inversion symmetry \mbox{${\bf r}\to-{\bf r}$} induces topological transport related to the Berry curvature, which in time-reversal invariant systems has odd parity, \mbox{${\bm \Omega}({\bf p}) = -{\bm \Omega}(-{\bf p})$}~\cite{xiao10,girvin19}. The Berry curvature impacts the semiclassical quasiparticle dynamics as~\cite{Sundaram1999,xiao10}
\begin{equation}
    \dot{\bf r} = {\bf v}({\bf p}) - \frac{e}{\hbar} \mathbfcal{E} (t) \times \boldsymbol{\Omega}({\bf p}) . \label{eq:velocity}
\end{equation}
The first term \mbox{${\bf v}({\bf p}) = \partial_{\bf p} \varepsilon({\bf p})$} is the band velocity and the second term is the anomalous velocity driven by the homogeneous  electric field $\mathbfcal{E}(t)$ of light with negligible magnetic component. 
In time-reversal symmetric systems, the Berry curvature plays a crucial role in nonlinear photocurrent generation~\cite{Belinicher_1978,Ivchenko_1978,Belinicher1980,Sipe_prb_2000,Takahiro_sciadv_2016,sodemann15,Rostami_prb_2018,Principi_prb_2019,Watanabe_prx_2021,Zhang_Fu_PNAS_2021,Rostami_prr_2020,Dai_Rappe_cpr_2023,Ma_Song_nrp_2023}. Specifically, nonlinear Hall~\cite{sodemann15,Du2021,toshio20,yamaguchi24,nakazawa24} and injection~\cite{Belinicher1980,Sipe_prb_2000,Takahiro_sciadv_2016,deJuan2017} photocurrents depend on the product of a relaxation time scale and an average over the Berry curvature. For example, the nonlinear Hall current, measured recently in 2D metals~\cite{Xu2018,Kang2019,Ma2019}, is determined by the Berry curvature dipole ${\bf D}=\langle\partial_{\bf p} \Omega \rangle = \sum_{\bf p} (\partial_{\bf p} \Omega) f_0$, which is an average over the Fermi-Dirac distribution function $f_0$. Like Ohm's law, however, this charge current is relaxed by momentum-nonconserving processes like impurity scattering and does not probe FL effects.

In the following, we explore the nonlinear thermoelectric response of topological FLs. This response is highly sensitive to electron interactions and depends on the topological heat capacitance expressed in terms of topological Berry curvature moments that we term the {\em Berry curvature capacity}:
\begin{equation} \label{eq:defCOmega}
C_\Omega = \frac{\partial \langle \Omega^2 \rangle}{\partial T} = \frac{1}{T} \sum_{\bf p} (\varepsilon-\mu) \Omega^2 (- \partial_\varepsilon f_0) 
\end{equation}
and the {\em velocity-curvature capacity}
\begin{equation} \label{eq:defCva}
C_{v_a} = \frac{\partial \langle v_a \Omega \rangle}{\partial T} = \frac{1}{T} \sum_{\bf p} v_a =(\varepsilon-\mu) \Omega (- \partial_\varepsilon f_0). 
\end{equation}
Both of these quantities have integrands with even parity and are thus finite in the presence of time-reversal symmetry. Intuitively, the form of these topological heat capacities is inferred by comparing with the conventional thermoelectric current \mbox{${\bf j}^{\nabla T} = \chi^{\nabla T} {\bm\nabla} T$} where the thermoelectric response function \mbox{$\chi^{\nabla T} \sim C_e \sim \partial_T\langle v^2 \rangle$} is proportional to the conventional electronic heat capacity~\footnote{To obtain the electronic heat capacity in correct units, consider $m^\ast_e C_e$, $m^\ast_e (e {\cal E}/\hbar)^2 C_\Omega$, and $m^\ast_e (e {\cal E}/\hbar) C_{v_a}$, where $m^\ast_e$ is the electronic effective mass.}. Incorporating the anomalous velocity into this definition naturally leads to the topological heat capacities characterized by Eqs.~\eqref{eq:defCOmega} and~\eqref{eq:defCva} with a respective quadratic and linear dependence on the electric field.

The topological capacities emerge in the nonlinear thermoelectric currents as sketched in Fig.~\ref{fig1}. We propose a second-order rectified (i.e., static) field-induced thermoelectric current that is proportional to $C_{\Omega}$. Specifically in the case of a circularly polarized field, this current is transverse to the direction of the temperature gradient [Fig.~\ref{fig1}(a)]. This nonlinear (in the electric fields) thermoelectric current takes the form~ \cite{suppl}:
\begin{equation}
\begin{split}\label{eq:jEEdT}
    &{\bf j}^0
  = T  \frac{e^3}{4 \hbar^2} \biggl( {\rm Im}[\tau_{\rm AN}(\omega)] [\mathbfcal{E}^T \times \mathbfcal{F}(\omega)]
 \\ &
   + {\rm Re}[\tau_{\rm AN}(\omega)] \Big \{ [\mathbfcal{E}^T\times \mathbfcal{E}(\omega)]_z [\mathbfcal{E}^\ast(\omega)\times \hat{\bf z}] + c.c.\Big\} 
 \biggr)C_{\Omega},
 \end{split}
 \end{equation}
where $\tau_{\rm AN}(\omega) = \tau_{\rm AN}/(1+i\omega \tau_{\rm AN})$, \mbox{$\mathbfcal{F}(\omega)= i \mathbfcal{E}(\omega)\times \mathbfcal{E}^\ast(\omega)$} is a real-valued circular-light induced field, and we introduce the effective thermal electric field \mbox{$\mathbfcal{E}^T= - {\bm \nabla T}/T$}~\cite{Luttinger1964,Tatara_PRL_2015,Nagaosa_prb_2020}. The first and second terms in Eq.~\eqref{eq:jEEdT} correspond to circular and linear polarized light responses, respectively. Crucially, for Fermi surfaces with hexagonal symmetry or higher, the anomalous velocity does not couple to a conserved current and has an enhanced odd-parity lifetime $\tau_{\rm AN}$, which in turn leads to a large increase in the magnitude of the current in Eq.~\eqref{eq:jEEdT}~\cite{suppl}. 

In addition, there is a dynamical field-driven thermoelectric current that is proportional to $C_{v_a}$ [Fig.~\ref{fig1}(b)],
\begin{equation}
\begin{split}
    {\bf j}^\omega
    &= - T  \frac{e^2}{\hbar} \biggl\{  \tau_{\rm FL} [{\bf C}_{\bf v} \cdot \mathbfcal{E}^T) (\mathbfcal{E}(\omega) \times \hat{\bf z}]
    \\ &
    \qquad -  \tau_{\rm AN}(\omega)  [\mathbfcal{E}^T \times \mathbfcal{E}(\omega)]_z {\bf C}_{\bf v} \biggr\}, \label{eq:jEdT}
    \end{split}
\end{equation}
which couples to both odd and even-parity relaxation times $\tau_{\rm AN}$ and $\tau_{\rm FL}$. The first term is a light-induced thermoelectric Hall current that is directed perpendicular to the polarization of the light field. The second term is proportional to the anomalous damping rate $\tau_{\rm AN}$ and thus enhanced in clean systems at low temperatures.

{\em  Model.} In the following, we derive the nonlinear currents in Eqs.~\eqref{eq:jEEdT} and~\eqref{eq:jEdT} using Boltzmann kinetic theory, where we consider transport in a single band. A general deformation of the quasiparticle distribution induces a nonzero expectation value of the charge current~\cite{son13,son13b}:
\begin{equation}
    {\bf j}(t, {\bf r}) = -e \sum_{\bf p}  \dot{\bf r} f(t,{\bf r},{\bf p}). \label{eq:current}
\end{equation}
By time-reversal symmetry, the velocity $\dot{\bf r}$ in Eq.~\eqref{eq:velocity} is odd under momentum inversion, and hence the current response is solely determined by odd-parity deformations. The time evolution of the quasielectron distribution follows from the FL kinetic equation~\cite{xiao05}:
\begin{equation}
    \biggl[\frac{\partial}{\partial t} + \dot{\bf r} \cdot \frac{\partial}{\partial {\bf r}} + \dot{\bf p} \cdot \frac{\partial}{\partial {\bf p}}\biggr] f(t,{\bf r},{\bf p}) = {\hat{\cal J}}\{\delta f\}  \label{eq:fullkinetic}
\end{equation} 
with \mbox{$\dot{\bf p} = -e \mathbfcal{E}(t)$}. 
The left-hand side is the streaming term that describes the free phase-space evolution in the presence of a Berry curvature, and the right-hand side is the collision integral that accounts for two-particle scattering. The collision integral will relax a general non-equilibrium distribution to the local equilibrium distribution \mbox{$f_0({\bf r},{\bf p}) = (\exp\{[\varepsilon({\bf p}) - \mu]/T({\bf r})\} + 1)^{-1}$}, which depends on a chemical potential $\mu$ and a local electronic temperature $T({\bf r})$ that varies in space. Using the chain rule, \mbox{$\partial_{\bf r} f_0 = - (\varepsilon-\mu) (-\partial_\varepsilon f_0) \mathbfcal{E}^T$}, we see that the temperature gradient \mbox{$\mathbfcal{E}^T$} induces a deviation from local equilibrium \mbox{$\delta f(t,{\bf r},{\bf p}) = f(t,{\bf r},{\bf p}) - f_0({\bf r},{\bf p})$}, which in turn generates a thermoelectric current.

We analyze the kinetic Eq.~\eqref{eq:fullkinetic} using a relaxation-time approximation. The leading deformation $\delta f$ solves~\cite{chen16,chen17}\cite{suppl}
\begin{equation}
 \label{eq:kineticf1}
\partial_t \delta f + {\hat{\cal J}}\{\delta f\} = e \mathbfcal{E} \cdot (\partial_{\bf p} f_0) - \Bigl[{\bf v} - \frac{e}{\hbar} \mathbfcal{E} \times \boldsymbol{\Omega}\Bigr] \cdot (\partial_{\bf r} f_0),
\end{equation}
which is sourced by the three terms on the right-hand side. Here, the first term describes the external Coulomb force, the second term is induced by the thermal gradient that couples to the band velocity, and the third term, which couples the temperature gradient to the anomalous velocity, is of topological origin. Note that, in 2D, the Berry curvature is normal to the 2D metal surface, \mbox{${\bm \Omega} ({\bf p}) = \Omega ({\bf p}) \hat{\bf z}$}, and the motion is always in plane. The relaxation-time approximation assumes that each source term $\psi$ is an eigenmode of the collision integral with a characteristic decay rate \mbox{${\hat{\cal J}}\{\psi\} = - \tau^{-1} \psi$}. For an interaction-dominated FL, these rates will in general be different: The Coulomb force induces a  shift in the center-of-mass momentum (as is seen by using the chain rule), which as discussed will not relax by binary collisions since they conserve the canonical momentum, and has thus infinite lifetime in a clean system. The second term in Eq.~\eqref{eq:kineticf1} induces a thermal energy current, which is a higher-order current mode with standard FL scaling \mbox{$\tau_{\rm FL} \sim 1/T^2$} (where we assume the interaction-dominated limit $\omega \tau_{\rm FL} \simeq 1$). The last term, however, describes the topological deformation. In anisotropic FLs, this mode does not induce a thermal current and will thus decay anomalously slowly with general \mbox{$\tau_{\rm AN}\sim 1/T^4$} (for which $\omega \tau_{\rm AN} \gg 1$ or, equivalently, $\tau_{\rm AN}/\tau_{\rm FL} \gg 1$).

Assuming a time-dependent electric field \mbox{${\cal E}_a(t) = {\rm Re} \{ {\cal E}_a(\omega) e^{i \omega t}\}$}, a solution of the kinetic Eq.~\eqref{eq:kineticf1} gives a deviation from equilibrium $\delta f = {\rm Re} \{\delta f^0 + \delta f^\omega e^{i\omega t} \}$ with~\cite{suppl}
\begin{equation} \label{eq:deltaf}
\begin{split}
\delta f^0 &= + \tau_{\rm FL} \, (\varepsilon-\mu) ({\bf v}\cdot \mathbfcal{E}^T) (-\partial_\varepsilon f_0)
,
\\[1ex]
\delta f^\omega &= - \frac{e}{i\omega} \, [{\bf v}\cdot\mathbfcal{E}(\omega)] (-\partial_\varepsilon f_0) 
  \\[1ex]
&
\quad - \frac{e}{\hbar} \tau_{\rm AN} \, (\omega) (\varepsilon-\mu) \{ [ \mathbfcal{E}(\omega) \times {\bm \Omega}] \cdot \mathbfcal{E}^T\}  (-\partial_\varepsilon f_0) .
\end{split}
\end{equation}
All of these contributions have odd parity. They induce a nonlinear current in Eq.~\eqref{eq:current} when combined with the anomalous velocity, and for the topological deformation (the second term in $\delta f^\omega$) also in combination with the band velocity. In principle, the leading deformation $\delta f$, which is linear in the position and momentum derivatives, acts as a source for a second-order correction $\delta f_2$ that is obtained from an iterative solution of the kinetic equation. We do not state this term since it has even parity, and accordingly, its contribution to the nonlinear current vanishes, as discussed after Eq.~\eqref{eq:current}. 
Note that the currents considered are nonlinear in the electric field $\mathbfcal{E}$ but linear in $\mathbfcal{E}^T$. Nonlinear temperature effects, such as nonlinear Nernst currents~\cite{zeng19}, are thus subleading in this counting.  
Note also that the third-order correction $\delta f_3$ has again odd parity and will give rise to a finite current. Since this contribution is of nontopological origin~\cite{suppl}, it will not involve an anomalously enhanced lifetime $\tau_{\rm AN}$. Moreover, this nontopological current is subleading in the hydrodynamic regime, where $\tau_{\rm FL}$ is the shortest time scale, because it is proportional to $\tau^3_{\rm FL}$ and thus can be neglected in comparison with currents linearly proportional to the electron lifetime.

{\em Anomalous field-induced thermoelectricity.} From Eq.~\eqref{eq:current} and using the solution in Eq.~\eqref{eq:deltaf}, the nonlinear charge current is expressed as
\begin{equation} \label{eq:ja}
j_a = \text{Re} \left\{ j_a^{0} + j_a^{\omega} e^{i\omega t} + j_a^{2\omega} e^{2i\omega t} \right\} ,
\end{equation}
where $j_a^{0}$ is the (rectified) direct current, $j_a^{\omega}$ corresponds to the first harmonic, and $j_a^{2\omega}$ denotes the second-harmonic generation. Neglecting the thermal gradient in Eq.~\eqref{eq:deltaf}, \mbox{$\mathbfcal{E}^T=0$}, we recover the Drude linear response and the known nonlinear Hall effect characterized by the Berry curvature dipole~\mbox{$D_{a}$}~\cite{sodemann15}. However, since this contribution is not sensitive to electron interactions, here, we focus on the interplay of the thermal gradient and the external electric field. 

We first discuss the rectified field-induced nonlinear thermoelectric current, which arises from the interplay of anomalous velocity and the second component of $\delta f^\omega$, which gives
\begin{equation} \label{eq:ja0}
  j_a^0 = \chi_{abcd}^{{\cal E E} \nabla T}  {\cal E}_b(\omega) {\cal E}^\ast_c(\omega)  {\cal E}^T_d ,
\end{equation}
with a response function 
proportional to $\tau_{\rm AN}$ and~$C_\Omega$:
\begin{equation} \label{eq:chi2}
 \chi_{abcd}^{{\cal E E} \nabla T} = T \frac{e^3}{2 \hbar^2} \epsilon_{acz} \epsilon_{bdz}  \tau_{\rm AN}(\omega)  C_{\Omega} ,
\end{equation}
where the Berry curvature capacity $C_{\Omega}$ is defined in Eq.~\eqref{eq:defCOmega}, $\epsilon_{abc}$ is the antisymmetric tensor, and we sum over repeated indices. The Berry curvature capacity also sets the magnitude of the second-harmonic field-induced thermoelectric current $j_a^{2\omega} = \chi_{abcd}^{{\cal E E} \nabla T}  {\cal E}_b(\omega) {\cal E}_c(\omega)  {\cal E}^T_d$. Note that there is a field-induced current that arises from the combination of a third-order deformation $\delta f_3$ (which again has odd parity) with the band velocity term. This correction, however, is not of topological origin and distinct from the response in Eq.~\eqref{eq:chi2} in that it is of third order in a relaxation time. To establish the experimental signatures of the field-induced thermoelectric current discussed in the introduction, we decompose the electric fields in Eq.~\eqref{eq:ja0} for circularly and linearly polarized light using \mbox{${\cal E}_b(\omega) {\cal E}^\ast_c(\omega) = {\rm Re} [{\cal E}_b(\omega) {\cal E}^\ast_c(\omega)] + \epsilon_{b c \ell}{\cal F}_\ell (\omega)/2i$}, which in coordinate-free form gives Eq.~\eqref{eq:jEEdT}. 

In addition to the rectified field-induced thermoelectric current, there is also a dynamical current:
\begin{equation}
    j_a^\omega = \chi_{abc}^{{\cal E} \nabla T} {\cal E}_b (\omega) {\cal E}^T_c ,
\end{equation}
which scales linearly with both electric field and thermal gradient. The corresponding response function is
\begin{equation}
\chi_{abc}^{{\cal E} \nabla T}  = - T \frac{e^2}{\hbar} \Big \{ \epsilon_{abz}\tau_{\rm FL} C_{v_c} - \epsilon_{cbz}  \tau_{\rm AN}(\omega) C_{v_a}\Big\}  ,
\end{equation}
where the velocity-curvature capacity $C_{v_a}$ is defined in Eq.~\eqref{eq:defCva}. The first contribution stems from $\delta f^0$ in combination with the anomalous velocity, and the second contribution arises from $\delta f^\omega$ combined with the band velocity. In coordinate-free form, we obtain Eq.~\eqref{eq:jEdT}. 

We proceed to establish the anomalous temperature scaling of the nonlinear response functions: Both the Berry curvature capacity and the velocity-curvature capacity are linear functions of temperature. Formally, this is seen using a Sommerfeld expansion~\cite{Ashcroft76}, which yields an angular integral over the Berry curvature:
 \begin{equation}\label{eq:QL}
    \begin{bmatrix}
      C_{\Omega} \\[5pt]   C_{v_a}  
    \end{bmatrix}
     \approx - \frac{\pi^2 T}{3} \frac{\partial}{\partial \mu}\int^{2\pi}_0 \frac{d\theta}{2\pi}  {\cal D}(\mu,\theta) 
     \begin{bmatrix}
       \Omega^2(\mu,\theta)
       \\[5pt]
         v_a (\mu,\theta) \Omega(\mu,\theta)
     \end{bmatrix} ,
\end{equation}
where ${\bf v}(\mu,\theta)$ is the Fermi velocity at a given angle $\theta$ and ${\cal D}(\mu,\theta)$ is the angle-resolved density of states. For FL lifetimes \mbox{$\tau_{\rm FL} \sim 1/T^2$}, this temperature dependence cancels to yield~\mbox{$\chi^{{\cal E}{\nabla T}}, \chi^{{\cal EE}{\nabla T}} \sim T^0$}. By contrast, an anomalous odd-parity lifetime \mbox{$\tau_{\rm AN} \sim 1/T^4$} gives a much larger magnitude that scales instead as \mbox{$\chi^{{\cal E} \nabla T}, \chi^{{\cal EE} \nabla T}\sim 1/T^2$}. This enhancement serves as a distinct and explicit signature of odd-parity dynamics. 

{\em Material realization.} In the remainder of this letter, we quantify the strength of field-induced thermoelectric effects by evaluating the parameters~$C_\Omega$ and~$C_{v_a}$ in a model system for topological insulator surface states such as Bi$_2$Te$_3$, which supports a protected Dirac fermion quasiparticle at the boundary~\cite{Zhang2009}. Our choice of material is motivated by the exceptional thermoelectric properties of bismuth chalcogenides, which exhibit the highest thermoelectric figure~\cite{WRIGHT1958,Pal2015,Witting2019,Fu_APL_2020,Wei2020,Balandin_nl_2010,Balandin_acs_2011}, and the recent observation of the nonlinear Hall effect in this bismuth-based compound~\cite{He2021,makushko2023tunable}. The thermoelectric properties and their interplay with electronic topology have attracted much theoretical interest~\cite{Balandin_prb_2010,Xu2017,Xu_PRL_2014,Baldomir2019}. Indeed, this system has a  hexagonal Fermi surface and a Berry curvature with threefold symmetry, which does not couple to the center-of-mass current and thus has anomalously large lifetime~$\tau_{\rm AN}$. 
\begin{figure}
    \centering
\includegraphics[width=8.5cm]{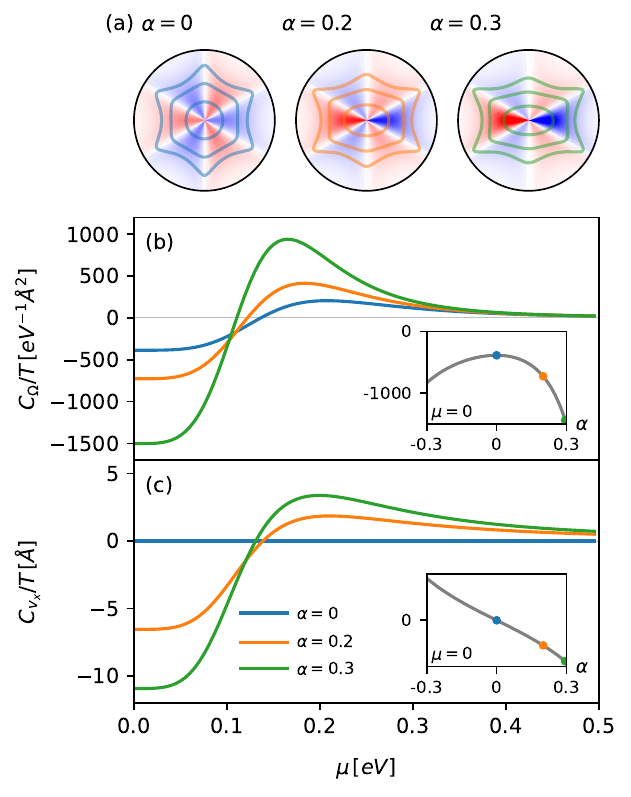}
    \caption{
    (a) Density plot of the Berry curvature~$\Omega({\bf p})$ for three strain parameters \mbox{$\alpha=0,0.2$}, and \mbox{$0.3$}. Continuous lines show the Fermi surface for three chemical potentials \mbox{$\mu=0.1,0.2$}, and \mbox{$0.3 \mathrm{eV}$}. The Berry curvature has odd parity and increases with increasing strain. (b) Berry curvature capacity $C_\Omega$ and (c) velocity-curvature capacity component $C_{v_x}$ as a function of chemical potential. Inset: Intrinsic values at \mbox{$\mu=0$} as a function of the strain parameters $\alpha$, where points mark values in the main plot.
    } 
    \label{fig:2}
\end{figure}

The surface states are described by a two-band effective Hamiltonian~\cite{Fu2009,Siu2016}:
\begin{equation}
\hat {\cal H} =\gamma_{x} p_x\hat\sigma_y- \gamma_{y} p_y \hat\sigma_x + \lambda (p^3_{x}-3p_{x}p^2_y)\hat\sigma_z, \label{eq:Bi2Te3H}
\end{equation}
where \mbox{$\gamma_{x}=(1-\alpha)v_F$} and \mbox{$\gamma_{y}=(1+\alpha)v_F$} with $v$ a Fermi velocity parameter, $\hat{\sigma}_{x,y,z}$ are spin Pauli matrices, and we include a phenomenological parameter~$\alpha$ to include uniaxial strain that breaks the full threefold rotational symmetry. Note that we have neglected a particle-hole asymmetry term which does not contribute to the Berry curvature. For numerical calculations, we use the Bi$_2$Te$_3$ parameters \mbox{$v=2.55\mathrm{eV\AA}$} and \mbox{$\lambda=250 \mathrm{eV\AA^3}$}~\cite{Fu2009}. The dispersion is 
\mbox{$\varepsilon ({\bf p})= \pm [v^2(\theta) p^2 + \lambda^2 p^6\cos^2(3\theta)]^{1/2}$}, with \mbox{$v(\theta) = v [1+\alpha^2-2\alpha\cos(2\theta)]^{1/2}$}, which has a hexagonal Fermi surface that is stretched in the $x$ direction for a positive strain parameter \footnote{Note that we consider the effect of strain on the Fermi velocity and do not include a possible shift of Dirac cone in the surface Brillouin zone.}, with a corresponding angle-resolved density of states ${\cal D}(\mu,\theta)  = \mu \{4\pi [v^2(\theta) + 3 \lambda^2 p^4_{\rm F}(\mu, \theta)\cos^2(3\theta)]\}^{-1}$, with $p_{\rm F}(\mu, \theta)$ being the Fermi wave vector. The Berry curvature reads 
$\Omega({\bf p}) = \Omega (\varepsilon,\theta) = \frac{\lambda (1-\alpha^2)}{v} \left(\frac{v p}{\varepsilon}\right)^3  \cos(3\theta)$, which is shown in Fig.~\ref{fig:2}(a) as a density plot in momentum space for three values of the strain parameter $\alpha=0,0.2,$ and $0.3$. 
The Berry curvature inherits the hexagonal symmetry but has odd parity, and it is enhanced by adding strain. We also include in each figure the Fermi surface contour for \mbox{$\mu=0.1,0.2$}, and \mbox{$0.3 \mathrm{eV}$}, with apparent hexagonal symmetry.

Figure~\ref{fig:2}(b) shows the leading temperature coefficient of the Berry curvature capacity $C_\Omega$, Eq.~\eqref{eq:QL}, as a function of the chemical potential $\mu$ for three values of the strain parameter \mbox{$\alpha=0,0.2$}, and $0.3$ (blue, orange, and green curves, respectively). Notably, at the Dirac point at small $\mu$, the Berry curvature is constant and the density of states vanishes as a linear function of $\mu$, such that $C_\Omega$ is finite with an intrinsic value $C_\Omega =- \pi \lambda^2 T/12v^4$
without strain. This is distinct from other topological transport coefficients like the Berry curvature dipole, which vanishes at the Dirac point~\cite{sodemann15}. The inset shows the intrinsic value as a function of the strain parameter, which is widely tunable. As the chemical potential increases, the Berry curvature capacity increases as well and changes sign, which is determined by an inflection point in the density of states along C$_{3v}$ symmetry axes.

Figure~\ref{fig:2}(c) shows the $x$ component $C_{v_x}$ of the velocity-curvature capacity for the same parameters as in Fig.~\ref{fig:2}(b). Quite generally, the mirror symmetry of the system forces~${\bf C}_v$ to vanish for any value $\lambda$ (blue curve), but the~$C_{v_x}$ component is nonzero if the threefold rotational symmetry is broken by strain, while~$C_{v_y}$ remains zero. In this regard, the symmetry requirement of the velocity-curvature capacity~${\bf C}_v$ is like that of the Berry curvature dipole~${\bf D}$~\cite{You_prb_2018}. The intrinsic plateau of~$C_{v_x}$ shows an almost linear dependence on the strain parameter $\alpha$ as depicted in the inset of Fig.~\ref{fig:2}(c). At finite chemical potential, the functional dependence of~$C_{v_x}$ is again linked to the density of states and like the Berry curvature capacity. These unique characteristics make these two quantities $C_\Omega$ and~${\bf C}_v$ promising for experimental investigations of the odd-parity relaxation dynamics in optically assisted thermoelectric spectroscopy.

To estimate the strength of the nonlinear light-induced electrothermal currents, we obtain the ratio:
\begin{equation}
    \frac{|j^{{\cal E}{\cal E}\nabla T}|}{|j^{\nabla T}|} = \frac{\alpha}{2\pi n}  \frac{\tau_{\rm AN}}{\tau_{\rm FL}} \frac{I}{I_0},
\end{equation}
which is valid near the intrinsic limit. Here, $\alpha$ is the fine-structure constant, $n$ is the refraction index, and $I$ is the intensity of incident light relative to the characteristic intensity \mbox{$I_0 = v^2/\hbar \lambda^2 = 1.65 \times 10^{13} {\rm Wm}^{-2}$}. For typical intensities \mbox{$I = 10^{11}--10^{13} {\rm W m}^{-2}$} and an enhancement \mbox{$\tau_{\rm AN}/\tau_{\rm FL} \approx 10^0--10^3$}, a thermoelectric current \mbox{$j^{\nabla T} = 1 {\rm mA}$} gives a nonlinear current \mbox{$j^{{\cal E}{\cal E}\nabla T} = 70{\rm nA}--0.7{\rm mA}$}, well within the range of current experiments.
To investigate these currents, the driving laser frequency should be smaller than the optical gap \mbox{$\hbar\omega < 2|\mu|$} to avoid photoelectron generation and minimize light-induced heating~\cite{Xu2018, Kang2019, Ma2019,He2021,makushko2023tunable}. 
In addition, an external magnetic field  breaks time-reversal invariance and decreases the nonlinear field-induced current, providing a further diagnostic tool of the anomalous dynamics.

\begin{acknowledgments}
We thank Alexander A. Balandin for helpful discussions. 
This letter is supported by Vetenskapsrådet with Grant No. 2020-04239 (J.H.) and Grant No. 2018-04252 (H.R.).
\end{acknowledgments}
 
\bibliography{references.bib} 

\end{document}


\title{Supplemental Material for ``Nonlinear thermoelectric probes of anomalous electron lifetimes in topological Fermi liquids"}

\author{Johannes Hofmann}
\email[Corresponding author: ]{johannes.hofmann@physics.gu.se}
\affiliation{Department of Physics, Gothenburg University, 41296 Gothenburg, Sweden}
\affiliation{Nordita, Stockholm University and KTH Royal Institute of Technology, 10691 Stockholm, Sweden}

\author{Habib Rostami}
\email[Corresponding author: ]{hr745@bath.ac.uk}
\affiliation{Department of Physics, University of Bath, Claverton Down, Bath BA2 7AY, United Kingdom}

\date{\today}
\maketitle
\tableofcontents

\section*{}
This supplemental material expands on the discussion in the main text and discuss the Fermi surface deformations to third order in the external forces for completeness. As discussed in the main text, the second-order nonlinear deformations do not contribute to the electrical current by parity, and third-order correction only participate for large temperature gradients. In addition, this supplemental material contains useful intermediate steps for the computation of the nonlinear thermoelectric current in terms of the topological heat capacitances, which are evaluated for a topological insulator with a hexagonally warped Fermi surface, for which we also give the anisotropic angle-resolved density of states.

\section{Kinetic equation}

The starting point is the kinetic equation for the Boltzmann distribution function $f$ discussed in the main text: 
\begin{align}
\partial_t f - e {\cal E}_a \partial_{p_a} f + \bigl[v_a - (\tfrac{e}{\hbar} \mathbfcal{E} \times \boldsymbol{\Omega})_a\bigr] \partial_{r_a} f &= {\hat{\cal J}}\{f-f_0\},
\end{align}
where $f_0$ is the local equilibrium distribution and ${\hat{\cal J}}\{\delta f\}$ represents the collision integral operator that acts on the deviation function $\delta f = f - f_0$. The deformation of the distribution function can be written in terms of odd and even harmonics at the Fermi surface, parameterized by the Fermi wave vector in polar coordinates, ${\bf p} = (p_F(\theta), \theta)$:
\begin{align}
   f_1 ({\bf p}) =\sum_{m\in \mathbb{Z}} \langle {\bf p}|\phi_m\rangle\langle \phi_m| f_1\rangle = \sum_{m \in \mathbb{Z}} f_{1,m} e^{im\theta}. 
\end{align}
where $|\phi_m\rangle$ is a harmonic basis function, which for a circular Fermi surface reads \mbox{$|\phi_m\rangle = e^{im\theta}$}. 
For the collision integral, we use a generalized relaxation time approximation, where deformations with odd and even parts (harmonics) relax at different rates, as discussed in the main text:
\begin{align}
 {\hat{\cal J}}|\delta f\rangle &= - \frac{1}{\tau_{\rm FL}} 
 \left\{ \sum_{m\in \text{even}}|\phi_m\rangle\langle \phi_m|\delta f\rangle - |\phi_0\rangle\langle \phi_0|\delta f\rangle  \right\}  -
  \frac{1}{\tau_{\rm AN}} 
 \left\{ \sum_{m\in \text{odd}}|\phi_m\rangle\langle \phi_m|\delta f\rangle - |\phi_\pm\rangle\langle \phi_\pm|\delta f\rangle  \right\}  
 \nonumber\\ &\quad -\frac{1}{\tau_{\rm FL}} |\phi_{\varepsilon,\pm}\rangle\langle \phi_{\varepsilon,\pm}|\delta f
\rangle .
\end{align}
Within this minimal relaxation-time model, all even harmonics (excluding \mbox{$m=0$}, the density zero mode) decay with the standard Fermi liquid rate $1/\tau_{\rm FL}$, while odd harmonics (excluding $m=\pm1$, the current zero mode) decay with anomalous rate $1/\tau_{\rm AN}$. Here, $|\phi_0\rangle$ and $|\phi_\pm\rangle$ are proportional to total charge and current densities respectively, which are conserved quantities the do not decay by binary collision. Moreover, as an exception, the higher harmonic basis function $|\phi_{\varepsilon,\pm}\rangle$,  with \mbox{$|\phi_{\varepsilon,\pm}\rangle = (\varepsilon-\mu) e^{\pm i\theta}$} for a circular Fermi surface, describes an energy current and decays with Fermi-liquid decay rate $\tau_{\rm FL}$ in spite of the fact that it has odd parity. A numerical calculation and perturbative analytical analysis of collision integral due to binary electron-electron interaction is given in Refs.~\cite{hofmann23,nilsson24}. Note that for an interaction-dominated system, we have the hierarchy of scales $\omega \tau_{\rm FL} \lesssim 1$, but  at the same time we have $\omega \tau_{\rm AN} \gg 1$.

\section{Iterative solution of the nonlinear response}

We model the light field as
\begin{align}
{\cal E}_a(t) &= {\rm Re} \{ {\cal E}_a(\omega) e^{i \omega t}\} = \frac{1}{2} \bigl( {\cal E}_a(\omega) e^{i \omega t} + {\cal E}_a^*(\omega) e^{-i \omega t}\bigr)
\end{align}
and expand the distribution function to third order ${\cal O}(\tau^3)$ in the collision rates,
\begin{align}
f(t) &= 
f_0 + \delta f_1 + \delta f_2 + \delta f_3 . 
\end{align}

To leading order, we obtain 
\begin{align}
    \partial_t \delta f_1 - {\hat{\cal J}}\{\delta f_1\} &= e {\cal E}_a(t) \partial_{p_a} f_{0} + (\tfrac{e}{\hbar} \mathbfcal{E}(t) \times \boldsymbol{\Omega})_a \partial_{r_a} f_{0}
    - v_a \partial_{r_a} f_{0} \label{eq:LO} ,
\end{align}
where the spatial derivative acting on the local distribution function gives a temperature derivative
\begin{align}
\partial_{r_a} f_0 &= (\nabla_a T) \frac{\partial f_0}{\partial T} = -(\varepsilon-\mu) \biggl(- \frac{\partial f_0}{\partial \varepsilon}\biggr) {\cal E}_a^T.
\end{align}
with ${\cal E}_a^T= -\nabla_a T/T$. Each external perturbation on the right-hand side of Eq.~\eqref{eq:LO} induces a separate deviation $f_1$ from local equilibrium that decays by collisions, where, as discussed above and in the main text, the rate of decay may be different for different perturbations.

\subsection{First-order correction}

Writing the first-order deformation in terms of static and dynamic components as $\delta f_{1} = {\rm Re} \{ \delta f_{1}^0 + \delta f_{1}^\omega e^{i\omega t} \}$ (where compared to the main text, we include the order as an additional superscript), we obtain
\begin{align}
\delta f_1^0 &= -\tau_{\rm FL} v_a (\partial_{r_a} f_0) \\[1ex]
\delta f_1^\omega &= (i\omega)^{-1} e {\cal E}_a(\omega) (\partial_{p_a} f_0) + \tau_{\rm AN}(\omega) (\tfrac{e}{\hbar} \mathbfcal{E}(\omega) \times \boldsymbol{\Omega})_a (\partial_{r_a} f_0) ,
\end{align}
where $\tau_{\rm AN}(\omega) = 1/(i \omega + \tau_{\rm AN}^{-1})$. All of these deformations have odd parity. 
Here, $\delta f_1^0$ is proportional to $|\phi_{\varepsilon,\pm}\rangle$ and thus decays with the standard Fermi liquid lifetime $\tau_{\rm FL}$. 
In addition, the first term in $\delta f_1^\omega$ is proportional to the current zero modes, which does not relax due to electron-electron interaction, hence $\tau=\infty$ for this mode, which thus has the prefactor $(i\omega)^{-1}$. On the other hand, the second term in $\delta f_1^\omega$ is induced by the anomalous velocity and not proportional to the current zero mode, and it will thus relax with $\tau_{\rm AN}(\omega)$.

\subsection{Second and third-order corrections}

Higher-order corrections follow from the recursive relation
\begin{align}
    \partial_t \delta f_{n+1} - {\hat{\cal J}}\{\delta f_{n+1}\} &= e {\cal E}_a{(t)} \partial_{p_a} \delta f_n - v_a  \partial_{r_a} \delta f_n
    + (\tfrac{e}{\hbar} \mathbfcal{E}{(t)} \times \boldsymbol{\Omega})_a \partial_{r_a} \delta f_n. \label{eq:orderiteration}
\end{align}
Note for further reference that the deformation $f_{n+1}$ derived in this way from $f_n$ is of one additional order in $\cal E$ (first term), ${\cal E}^T$ (second term), or even ${\cal E} {\cal E}^T$ (third term). Higher-order corrections are formally expressed as $\delta f_{2} = {\rm Re} \{ \delta f_{1}^0 + \delta f_{1}^\omega e^{i\omega t} + \delta f_{2}^{2\omega} e^{2i\omega t} \}$ and written in concise notation as:
\begin{align}
\delta f_2^0 &= {\hat{\cal J}}^{-1} \left\{ e {\cal E}^\ast_a(\omega) \partial_{p_a} \delta f_1^\omega + (\tfrac{e}{\hbar} \mathbfcal{E}^\ast(\omega) \times \boldsymbol{\Omega})_a  \partial_{r_a} \delta f_1^\omega
- v_a  \partial_{r_a} \delta f_1^0\right\} 
\\[1ex]
\delta f_2^\omega &= [i\omega+{\hat{\cal J}}]^{-1} \left\{ e {\cal E}_a(\omega) \partial_{p_a} \delta f_1^0 + (\tfrac{e}{\hbar} \mathbfcal{E}(\omega) \times \boldsymbol{\Omega})_a  \partial_{r_a} \delta f_1^0
- v_a  \partial_{r_a} \delta f_1^\omega\right\}
\\[1ex]
\delta f_2^{2\omega} &= [2i\omega+{\hat{\cal J}}]^{-1}  \left\{ e {\cal E}_a(\omega) \partial_{p_a} \delta f_1^\omega + (\tfrac{e}{\hbar} \mathbfcal{E}(\omega) \times \boldsymbol{\Omega})_a  \partial_{r_a} \delta f_1^\omega
- v_a  \partial_{r_a} \delta f_1^{2\omega}\right\} .
\end{align}
Since $\delta f_1$ is an odd function of $\bf p$,  $\delta f_2$ is an even function of $\bf p$ and will thus does not contribute to current. \\

The third-order correction $\delta f_3$, however, is an odd function of $\mathbf{p}$, and therefore, in principle, can contribute. Since we study the response at quadratic order in the electric field and linear order in the temperature gradient, the Berry-curvature term (the third term in Eq.~\eqref{eq:orderiteration}) will not contribute at this order.  Moreover, since $\delta f_3$ is at least of third order, it will only contribute to the current through the band velocity term. The third-order correction $\delta f_3$ and its current are thus of non-topological origin. Since the only higher-order odd-parity deformation with relaxation time $\tau_{\rm AN}$ is linked to the Berry curvature term, it will not contribute to the current induced by the third-order correction, which is thus of order ${\it O}(\tau_{\rm FL} (i\omega)^{-2})$, ${\it O}(\tau_{\rm FL}^2 (i\omega)^{-1})$, or ${\it O}(\tau_{\rm FL}^3)$, which are equivalent in our counting with $\omega \tau_{\rm FL} \simeq 1$. Compared to the topological currents obtained from $\delta f_1$ described in the main text, these non-topological third-order currents are subleading in the hydrodynamic regime, where $\tau_{\rm FL}$ is the shortest timescale, and the term proportional to $\tau^3_{\rm FL}$ can be neglected in comparison to those linear in $\tau_{\rm FL}$ and $\tau_{\rm AN}$.

\section{Thermometric current} 

We are interested in the rectified and first harmonic light-induced thermoelectric current, which we write as
\begin{align}
{\bf j} &= {\rm Re}\bigl\{{\bf j}^0 + {\bf j}^\omega e^{i\omega t} + {\bf j}^{2\omega} e^{2i\omega t}\bigr\} \nonumber \\
 &= - e \sum_{\bf p} {\bf v} \, {\rm Re} \bigl\{\delta f_1^0 + \delta f_1^\omega e^{i\omega t}\bigr\} + e \sum_{\bf p} \biggl[\frac{\tfrac{e}{\hbar} \mathbfcal{E}(\omega) e^{i \omega t} + \tfrac{e}{\hbar} \mathbfcal{E}^*(\omega) e^{-i \omega t}}{2} \times \boldsymbol{\Omega}\biggr] \, {\rm Re} \bigl\{\delta f_1^0 + \delta f_1^\omega e^{i\omega t}\bigr\} ,
\end{align}
such that
\begin{align}
{\bf j}^0 &= - e \sum_{\bf p} {\bf v} \, \delta f_1^0 + \frac{e}{\hbar} \sum_{\bf p} \biggl[\frac{1}{4} (e \mathbfcal{E}^*(\omega) \times \boldsymbol{\Omega}) \delta  f_1^\omega + \frac{1}{4} (e \mathbfcal{E}(\omega) \times \boldsymbol{\Omega}) \delta f_1^{\omega*}\biggr] \\[1ex]
&= - e \sum_{\bf p} {\bf v} \, \delta f_1^0 + \frac{e^2}{2 \hbar} {\rm Re} \sum_{\bf p} (\mathbfcal{E}^*(\omega) \times \boldsymbol{\Omega}) \delta f_1^\omega 
\end{align}
and
\begin{align}
{\bf j}^\omega &= - e \sum_{\bf p} {\bf v} \, \delta f_1^\omega + \frac{e^2}{\hbar} \sum_{\bf p} (\mathbfcal{E}(\omega) \times \boldsymbol{\Omega}) \, \delta f_1^0 .
\end{align}
Evaluating these expression using the result for the quasiparticle deformation $f_1$ gives the expressions stated in Eqs.~(11) and~(13) of the main text.

\subsection{Sommerfeld expansion}

The Sommerfeld expansion reads
\begin{align}
\int_{-\infty}^\infty d\varepsilon \, H(\varepsilon) (- \partial_\varepsilon f_0) &= H(\mu) + \frac{\pi^2}{6} T^2 \partial^2_\mu H(\mu) + \ldots
\end{align}
Rewrite the momentum integration as
\begin{align}
\sum_{\bf p} \ldots &= \int_{-\infty}^\infty d\varepsilon \sum_{\bf p} \delta(\varepsilon - \varepsilon({\bf p})) \ldots = \int_0^{2\pi} \frac{d\theta}{2\pi} \int_{-\infty}^\infty d\varepsilon \, D(\varepsilon, \theta) \ldots ,
\end{align}
where $D(\varepsilon, \theta)$ is the angle-dependent density of states. We compute the Fermi-surface average $(\varepsilon-\mu) F(\varepsilon,\theta)$ as follows: 
\begin{align}
\sum_{\bf p} (\varepsilon-\mu) F(\varepsilon,\theta) (- \partial_\varepsilon f_0) 
&= \int_0^{2\pi} \frac{d\theta}{2\pi} \int_{-\infty}^\infty d\varepsilon \, (\varepsilon-\mu) F(\varepsilon,\theta) (- \partial_\varepsilon f_0) \nonumber \\
&= \frac{\pi^2}{6} T^2 \int_0^{2\pi} \frac{d\theta}{2\pi} \, \Bigl[D(\varepsilon, \theta) (\varepsilon-\mu) F(\varepsilon,\theta)\Bigr]^{''}_{p=p_F(\mu,\theta)} \nonumber \\
&= \frac{\pi^2}{3} T^2 \int_0^{2\pi} \frac{d\theta}{2\pi} \, \Bigl[D(\varepsilon, \theta) F(\varepsilon,\theta)\Bigr]^{'}_{p=p_F(\mu,\theta)} .
\end{align}
Defining $F=\Omega^2$ and $F=v_a \Omega$, we obtain the relation for topological heat capacitance given in Eq. (15) of the main text.  
 
\subsection{Angle-resolved density of states in topological insulators}

For an anisotropic Fermi surface, the Fermi momentum will depends on the polar angle $\theta$ of the momentum, i.e. $p_F(\mu,\theta)$. Therefore, we define an angle resolved density of states 
\begin{align}
D(\varepsilon, \theta) &= \frac{1}{2\pi} \int_0^\infty dp \, p \, \delta(\varepsilon - \varepsilon({\bf p})) 
= \frac{1}{2\pi} \frac{p_F(\mu,\theta)}{\frac{\partial \varepsilon({\bf p})}{\partial p} \bigr|_{p=p_F(\mu,\theta)}} .
\end{align}
We include the conduction band energy dispersion of the topological insulator with hexagonal warping by diagonalizing the Hamiltonian described by Eq. (16) of the main text. It explicitly reads: 
\begin{equation}
 \varepsilon_{\bf p} = \sqrt{v^2(\theta) p^2+ \lambda^2 p^6 \cos^2(3\theta)} ,
\end{equation}
which gives
\begin{equation}
  \frac{\partial  \varepsilon_{\bf p}}{\partial p}\Big|_{p= p_{\rm F}(\mu,\theta)} = \frac{2v^2(\theta) p + 6 \lambda^2 p^4\cos^2(3\theta)}{ \sqrt{v^2(\theta) p^2+ \lambda^2 p^6 \cos^2(3\theta)}}  \Bigg|_{p= p_{\rm F}(\mu,\theta)} = p_{\rm F}(\mu,\theta)
  \frac{2v^2(\theta) + 6 \lambda^2 p^4_{\rm F}(\mu,\theta)\cos^2(3\theta)}{\mu} ,
\end{equation}
and
\begin{equation}
   {\cal D}(\mu,\theta) = \frac{1}{2\pi} \frac{p_{\rm F}(\mu,\theta)}{  \frac{\partial  \varepsilon_{\bf p}}{\partial p}\Big|_{p= p_{\rm F}(\mu,\theta)}} = \frac{1}{4\pi} \frac{\mu}{v^2(\theta) + 3 \lambda^2 p^4_{\rm F}(\mu,\theta)\cos^2(3\theta)}, 
\end{equation}
where the Fermi momentum follows the following relation 
\begin{equation}
\mu^2 = v^2(\theta) p^2_{\rm F}(\mu,\theta)+ \lambda^2 p^6_{\rm F}(\mu,\theta) \cos^2(3\theta). 
\end{equation}
This relation can be analytically solved, which determines $p_{\rm F}$ as a function of polar angle $\theta$ at a given chemical potential $\mu$. Finally, the total density of state reads 
\begin{equation}
D(\mu) = \int^{2\pi}_0\frac{d\theta}{2\pi} {\cal D}(\mu,\theta) = \int^{2\pi}_0 \frac{d\theta}{8\pi^2} 
\frac{\mu}{v^2(\theta) + 3 \lambda^2 p^4_{\rm F}(\mu,\theta)\cos^2(3\theta)}.
\end{equation}

\bibliography{references}